\documentclass[preprint,5p,times,sort&compress]{elsarticle}
\pdfoutput=1
\usepackage{graphicx}  
\usepackage{ifthen}
\usepackage{amssymb}
\usepackage{wasysym}
\usepackage{textcomp}
\usepackage{ifthen}
 \usepackage{lineno}
 \usepackage{siunitx}
\DeclareSIUnit{\sample}{S}
\DeclareSIUnit{\bit}{bit}
\DeclareSIUnit{\byte}{Byte}
\usepackage{url}

\newboolean{parabolic}
\setboolean{parabolic}{false}
\journal{Nuclear Instruments and Methods in Physics Research, A}

\begin{document}
\begin{frontmatter}

\title{Performance Tests of Feature Extraction Algorithms for Short Preamplifier Transients}
\author[1]{Holger~Flemming\corref{cor}}
\ead{h.flemming@gsi.de}

\author[2,3]{Oliver~Noll}
\ead{ernoll@uni-mainz.de}

\cortext[cor]{Corresponding author}
\affiliation[1]{organization={GSI Helmholtzzentrum f{\"u}r Schwerionenforschung GmbH},
addressline={Planckstra{\ss}e 1},
postcode={64291},
city={Darmstadt},
country={Germany}}
\affiliation[2]{organization={Johannes Gutenberg-Universität Mainz},
postcode={55099},
city={Mainz},
country={Germany}}
\affiliation[3]{organization={Helmholtz Institut Mainz},
addressline={Staudingerweg 18},
postcode={55128},
city={Mainz},
country={Germany}}


%
%


\begin{abstract}
In this work the estimation of pulse time and amplitude from short pulse transients is analysed as they are recorded by a transient recorder ASIC called ATR16. Algorithms of different complexity are tested with simulated preamplifier data processed with a VHDL model of the ATR16 ASIC. It is shown that a time precision below \SI{1}{\nano\second} and a signal amplitude precision below \SI{1}{\mega\electronvolt} energy equivalent is feasible.
\end{abstract}



\end{frontmatter}

\section{Introduction}
While the baseline approach for the PANDA EMC readout\cite{emc-tdr} foresees a transmission of the analogue signals produced by the preamplifier and shaper ASIC called APFEL\cite{irrDB114} over long cables to outside sampling ADCs a new concept is based on digitiser ASICs placed close to the preamplifier. This ASIC called ATR16 is an analogue transient recorder\cite{twepp21-TR}. 

\begin{figure}[htb]
\includegraphics[width=\linewidth]{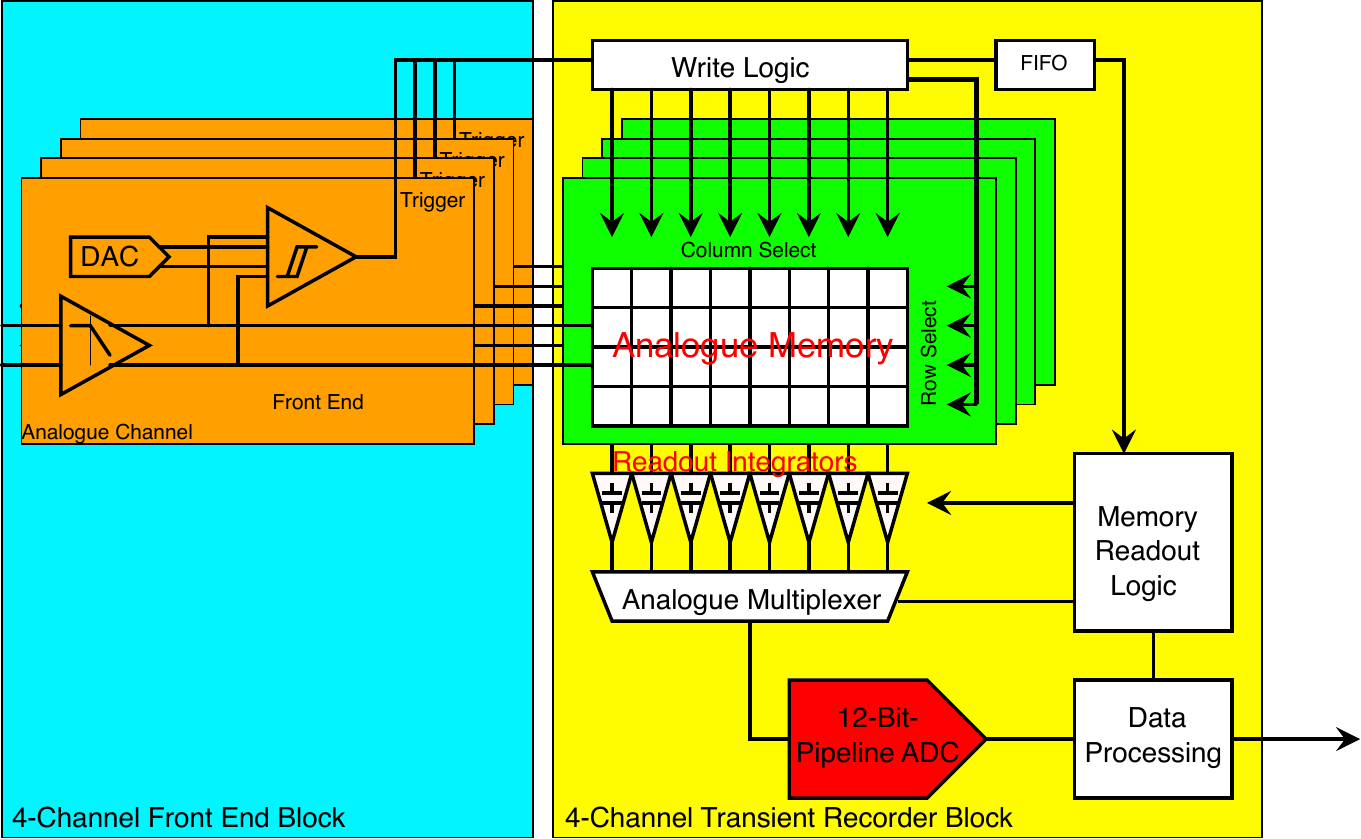}
\caption{\label{pic:TranRec}Block diagram of the front end and transient recorder unit of the ATR16.}
\end{figure}

Figure \ref{pic:TranRec} shows a block diagram of the transient recorder unit of the ATR16. Each recorder block is connected to a four-channel front end. Main component of the recorder unit is an analogue memory. The incoming analogue signal is sampled with up to \SI{100}{\mega \sample\per\second} and written into the analogue memory in a cyclic manner. When a pulse is detected the write logic switches to the next row after a configurable delay. This way the complete pulse transient is stored in the memory and can be read out asynchronously by the readout circuit consisting of charge integrators and an analogue multiplexer. Afterwards the transient is digitised with a \SI{33}{\mega\sample\per\second}, 12-bit pipeline ADC which is shared by the four channels.\\

In the past several groups studied the feature extraction of detector signals digitised with sampling ADCs in continuous operation\cite{irrDB113,irrDB207,OliverNoll}. Here elaborated digital filter and estimation algorithms can be utilised which is not possible with short finite traces as provided by the ATR16. So this work looks into the feature extraction of such short traces and the resulting contributions to energy and time precision which are required to be less than \SI{1}{\mega\electronvolt} and \SI{3}{ns} respectively\cite{emc-tdr}.\\

In section \ref{sec:architecture} the optimum sampling frequency and trace length is derived from the preamplifier signal characteristics. Then the examined algorithms for time and amplitude estimation are described in section \ref{sec:featextr}.  In section \ref{sec:sim} the simulation environment is described and the results are presented in section \ref{sec:results} with a final outcome concluded in section \ref{sec:conclusion}.

\section{Optimisation of Sampling Parameter}\label{sec:architecture}
Due to the architecture of the ATR16 which foresees an analogue transient recording, each storage cell has to be realised by a capacitor, implementation of which is area consuming. So to minimise the number of samples in a single trace the sampling frequency should be close to the physical limit. From Shannon's first theorem\cite{irrDB42} it is known that the signal is completely defined when the sampling frequency is the doubled maximum frequency of the signal.\\

Following Sansen and Chang\cite{irrDB49} the spectrum of the output signal of a charge sensitive amplifier is given by

\begin{equation}
H(s) = \frac{q}{sC_f} \left[ \frac{s\tau_0}{1 + s\tau_0} \right] \left[ \frac{A}{1+s\tau_0} \right]^n 
\end{equation}

corresponding to a pulse function of a pulse starting at $t_0$ in the time domain given by

\begin{equation}
V_{out}(t) = \frac{qA^nn^n}{C_f n!} \left( \frac{t-t_0}{\tau_s} \right)^n e^{-n(t-t_0)/\tau_s}\label{eq_pasa_pulse}
\end{equation}

with the incoming charge $q$, the stage amplification $A$, the feedback capacitance $C_f$, the shaper order $n$, the time constant of a single integrator stage $\tau_0$ and the shaping time $\tau_s = n\tau_0$.\\

\begin{figure}[tb]
\includegraphics[width=\linewidth]{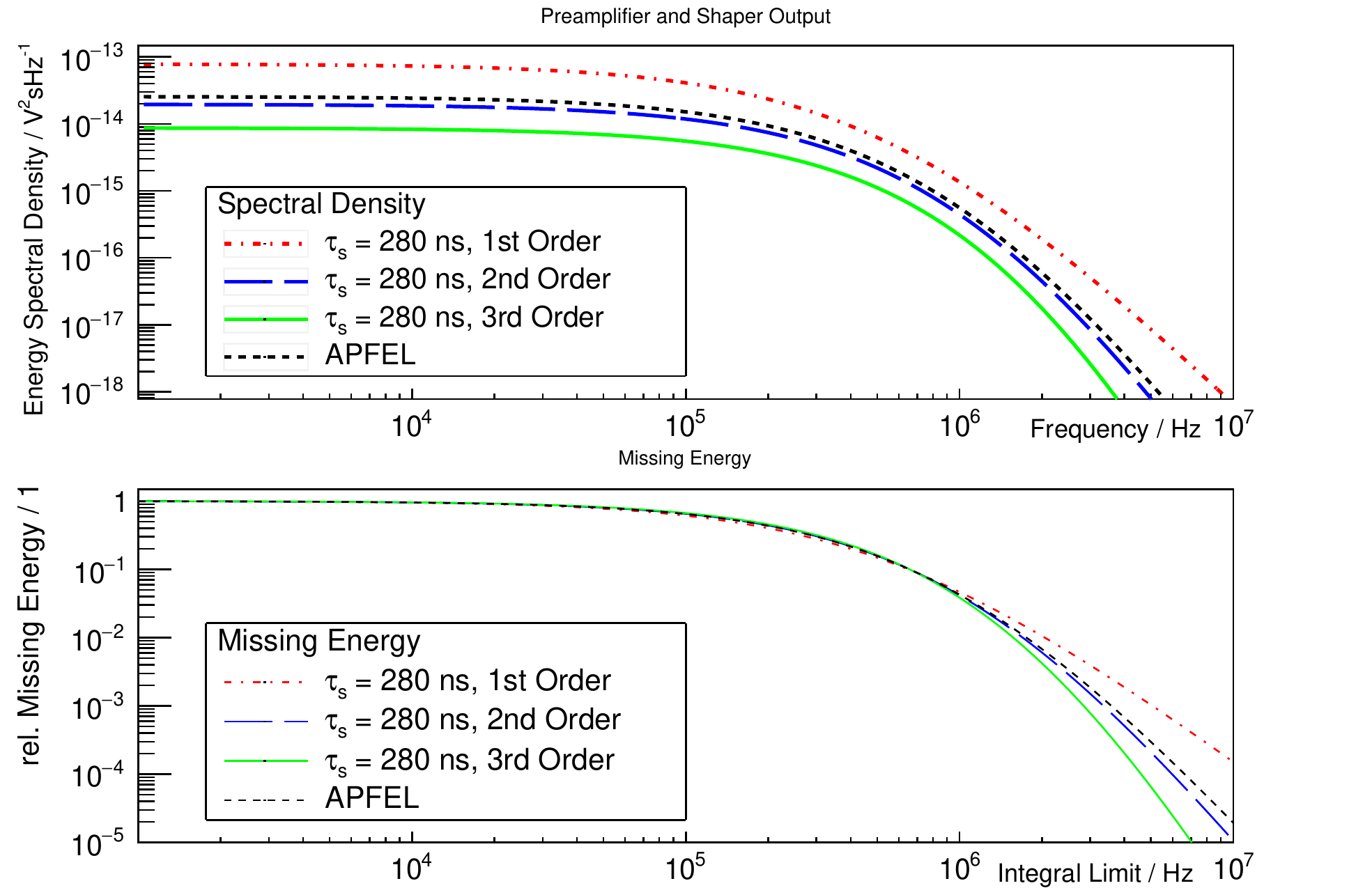}
\caption{\label{pic:EnSpec} Energy spectral density of a charge sensitive amplifier signal with a CR - RC$^n$ shaper for $q=\SI{1}{\pico\coulomb}$, $C_f = \SI{1}{\pico\farad}$ and $A=1$ (top) and the relative missing energy for integrated spectral densities (bottom).}
\end{figure}

From fitting analysis of the analytical pulse function in equation \ref{eq_pasa_pulse} to measured pulses of the APFEL ASIC it came out that the pulses are described best with a shaping time of $\tau_s = \SI{280}{\nano\second}$ and an order of $n = 1.75$. In figure \ref{pic:EnSpec} the corresponding energy spectral densities for a shaping time of \SI{280}{\nano\second} and shaper orders of 1,2,3 and 1.75 are shown.\\

As the signals have a continuous spectrum up to infinite frequencies one has to define an arbitrary limit given by the fraction of the spectral power below a certain frequency to obtain a criterion for a reasonable sampling frequency. So the spectral energy density was integrated and the relative missing energy is drawn in the lower plot of figure \ref{pic:EnSpec}. One can find for the APFEL ASIC (black line) that 99.9~\% of the spectral energy is in the frequency range below \SI{3.62}{\mega\hertz} which leads to a minimum sampling frequency of approximately \SI{7.2}{\mega\hertz}. For this work a sampling frequency of \SI{8}{\mega\hertz} was chosen.\\

As the full pulse length is in the order of 1~\textmu s 8 samples are required to cover the full pulse. To be able to estimate the baseline from the transient, a number of 16 samples was chosen for transient recording. Accordingly the analogue memory shown in figure \ref{pic:TranRec} is organised in 16 columns.

\section{Feature Extraction Algorithms}\label{sec:featextr}
For this analysis several algorithms for estimation of pulse time and pulse amplitude are evaluated. A common requirement for all of these algorithms is that they might be implemented in hardware i.e. on the ATR16 ASIC itself or on an FPGA.\\

\begin{figure}[htb]
\includegraphics[width=\linewidth]{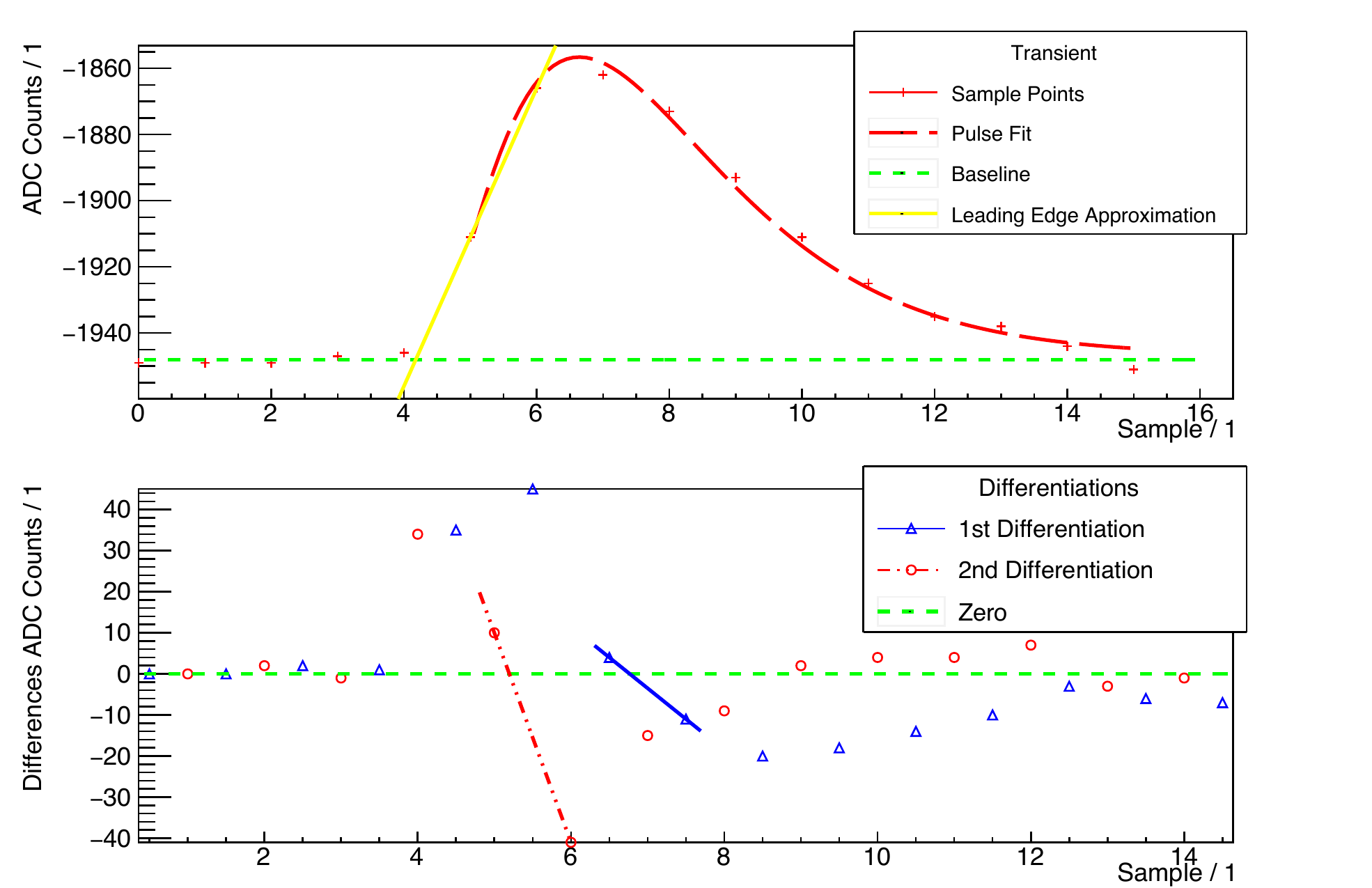}
\caption{\label{pic:timeExtr}Typical 16 sample trace of a preamplifier and shaper output pulse with estimated baseline and leading edge approximation and numerical differentiations.}
\end{figure}

Figure \ref{pic:timeExtr} shows a typical trace.  The recorded traces consist of 16 samples $a_i$ with $0 \le i \le 15$ which are shown in the upper plot of figure \ref{pic:timeExtr} as red marks. The dashed red line is a fit of the analytic function of the signal. In the lower plot the differentiated signal $d_i^1 = a_{i+1} - a_i$ with $0 \le i \le 14$ is shown as blue marks and the twice differentiated signal $d_i^2 = d_{i+1}^1 - d_i^1$ with $0 \le i \le 13$ is shown as red marks.\\

For both the time estimation and the amplitude estimation in common the baseline has to be known. Therefore a threshold $D_{th}$ is used to define the beginning of the pulse. So the index of the first sample of the pulse $i_{tr}$ is given by the first difference $d_i$ that fulfils $d_{i_{tr}}^1 >  D_{th}$. All samples before this point are taken as baseline. So the baseline level $\bar{b}$ is given by

\begin{equation}
\bar{b} = \frac{1}{i_{tr}} \sum_{i = 0}^{i_{tr}-1} a_i
\end{equation}

The baseline is shown as a green line in the upper plot of figure \ref{pic:timeExtr}. In the following, the time and amplitude estimation algorithms are described in detail.

\subsection{Time Estimation}
The aim of the time estimation is to determine the pulse time with a better granularity than given by the sampling frequency. The fraction of a sampling interval is called fine time. It is a measure of the pulse phase in relation to the sampling frequency.

\subsubsection{Baseline Crossing}
The idea of the baseline crossing algorithm is to place a tangent on the leading edge of the pulse and calculate the crossing point with the baseline. Therefore $d_{i_I}^1$ with $i_I > i_{tr}$ and $d_{i_I}^1 >  d_{i_I+1}^1$ is searched.\\

$a(t) = d_{i_I}^1 t + a_{i_I}$ describes the tangent at $i_I$ shown as yellow line in figure \ref{pic:timeExtr}. For $t_0$ at the baseline crossing $\bar{b} = a(t_0)$ one gets

\begin{equation}
t_0 = \frac{\bar{b} - a_{i_I}}{d_{i_I}^1}
\end{equation}

\subsubsection{Pulse Maximum}\label{sec:TimePulseMax}
At the pulse maximum the first derivative has to be zero. Therefore the time of the pulse maximum can be determined by calculating the zero crossing point between $d_{i_m}^1$ with $d_{i_m}^1 \ge 0$ and $d_{i_m+1}^1$ with $d_{i_m+1}^1 < 0$. The connecting line between $d_{i_m}^1$ and $d_{i_m+1}^1$ (shown as a blue line in figure \ref{pic:timeExtr}) is given by $d^1(t) = \alpha t + d_{i_m}^1$. For $d^1(t) = 0$ one gets

\begin{equation}
t_0 = \frac{d_{i_m}^1}{d_{i_m}^1-d_{i_m+1}^1}
\end{equation}

\subsubsection{Inflexion Point}
In the same way, the time of the pulse maximum can be calculated by the zero crossing of $d_i^1$ the time of the inflexion point on the leading edge can be determined by the zero crossing of $d_i^2$. When $d_{i_{In}}^2$ with $d_{i_{In}}^2 \ge 0$ and $d_{i_{In}+1}^2$ with $d_{i_{In}+1}^2 < 0$ one gets 

\begin{equation}
t_0 = \frac{d_{i_{In}}^2}{d_{i_{In}}^2-d_{i_{In}+1}^2}
\end{equation}

This interpolation is shown as red line in the lower plot of figure \ref{pic:timeExtr}.

\ifthenelse{\boolean{parabolic}}{%
\subsubsection{Parabolic Approximation}\label{sec:par_approx}
This algorithm uses the pulse maximum $a_m$ and the adjacent samples $a_{m-1}$ and $a_{m+1}$ for a parabolic interpolation of the pulse maximum. Let

\begin{equation}
p(t) = \alpha_2 t^2 + \alpha_1 t + \alpha_0\label{eq_par_approx}
\end{equation}

 with $a(-1) = a_{m-1}$, $a(0) = a_m $ and $a(1) = a_{m+1}$. By solving these equations one gets 
 
 \begin{eqnarray}
 \alpha_0 &=& a_m\\
 \alpha_1 &=& \frac{1}{2} ( a_{m+1} - a_{m-1})\\
 \alpha_2 &=& \frac{1}{2} ( a_{m+1} + a_{m-1} - 2 a_m )
 \end{eqnarray}
 
Finally one can calculate the position $t_m$ of the maximum of this parabolic interpolation by

\begin{equation}
t_m = \frac{a_{m+1} - a_{m-1}}{2(a_{m+1} + a_{m-1} - 2a_m)}\label{eq_time_par_approx}
\end{equation}
}{}

\subsection{Amplitude Estimation}\label{sec:ampExtr}
As well as for the time estimation for the amplitude estimation a series of different algorithms has been tested. Beginning with a simple maximum value estimation, integral calculation of the whole pulse as well as for defined windows \ifthenelse{\boolean{parabolic}}{to a parabolic interpolation of the pulse maximum}{} and a linear regression of the recorded transient and a standard pulse shape. Below the different algorithms are discussed in detail.\\

\subsubsection{Pulse Maximum}\label{sec:AmpPulseMax}
This is the simplest possible algorithm. The highest sample $a_m$ of the transient  is determined. The signal amplitude is the difference between this sample and the baseline.

\begin{equation}
S_{pm} = a_m - \bar{b}
\end{equation}

\subsubsection{Pulse Integral}

As explained above a threshold is used to define the beginning of pulse and to separate baseline samples from pulse samples. This algorithm accumulates the difference of all pulse samples  and the baseline to calculate a pulse integral. 

\begin{equation}
S_{Int} = \sum_{i=i_{tr}+1}^{15} (a_i - \bar{b})
\end{equation}

\subsubsection{Window Integral}

Unlike to the previous algorithm the window integral algorithm only accumulates a defined number of samples around the pulse maximum. Therefore, the position of the pulse maximum $m$ is determined. Afterwards, the window is defined by the number of samples before the maximum $n_b$ and the number of samples after the maximum $n_a$. Then the signal amplitude is given by

\begin{equation}
S_{WInt} = \sum_{i = m-n_b}^{m+n_a} (a_i - \bar{b})
\end{equation}

In this work two window definitions are analysed. The first one is $n_b = 1$ and $n_a = 2$ and the second one is $n_b = 2$ and $n_a = 4$. One should take note that the pulse maximum algorithm described in section \ref{sec:AmpPulseMax} could be considered as a special case of the window integral algorithm with $n_b = n_a = 0$.

\ifthenelse{\boolean{parabolic}}{%
\subsubsection{Parabolic Interpolation of Pulse Maximum}
The parabolic approximation of the pulse peak was already described for time estimation in section \ref{sec:par_approx}. With the time of the pulse maximum $t_m$ from equation \ref{eq_time_par_approx} one can calculate the maximum point $a(t_m)$ with equation \ref{eq_par_approx}. The pulse amplitude is given by 

\begin{equation}
S_{PA} = p(t_m) - \bar{b}
\end{equation}
}{}
\subsubsection{Linear Regression of Transient and Standard Pulse Shape}\label{sec:linregression}
The last algorithm uses the fact that the pulse shape of the pulses generated by the APFEL ASIC is well known. The pulse shape of a charge sensitive amplifier was already given in equation \ref{eq_pasa_pulse}. We get $t_0$ from the time estimation and calculate the samples of a normalised pulse

\begin{equation}
V_i = \left( \frac{iT - t_0}{\tau_s} \right)^n e^{-n(iT - t_0)/\tau_s}
\end{equation}

with the sampling interval $T$. The measured sample $a_i$ and $V_i$ are connected by 

\begin{equation}
a_i = S_{LR} V_i + b + \delta_i
\end{equation}
 with the signal amplitude $S_{LR}$, the baseline $b$ and an error $\delta_i$ which is caused by noise for example. To calculate $S_{LR}$ and $b$ from the samples $a_i$ a linear regression is used. \\

\section{Simulation}\label{sec:sim}
The described time and signal amplitude estimation algorithms were tested by simulations. For that a VHDL simulation of the ATR16 ASIC including modelling of the analogue components was used. The numerical stimulus is based on noise and signal analysis of the APFEL ASIC\cite{OliverNoll}.

\subsection{Stimulus Generation}
In section \ref{sec:architecture} the analytic expression of an APFEL pulse was introduced in equation \ref{eq_pasa_pulse}. Using Stirling's approximation $A^n n^n / C_f n!$ can be expressed as $G_{PASA} \cdot e^n$ with the preamplifier gain $G_{PASA}$. For the PANDA EMC the input charge $q$ is given by 
\begin{equation}
q = G_{APD}  \cdot Y \cdot A_{rel} \cdot Q_{eff} \cdot e
\end{equation}

with the APD gain $G_{APD}$, the light yield of the lead tungstate crystals $Y$, the relative area coverage of the APDs $A_{rel}$, the quantum efficiency of the APDs $Q_{eff}$ and the elementary charge $e$. The values of these parameters used in the simulation are given in table \ref{tab:APFELParameter}. The pulse shape characteristics $n$ and $\tau_s$ are fixed by the analysis of measured APFEL pulses.\\

\begin{table}[htb]
\caption{\label{tab:APFELParameter}Parameter used for simulation stimulus generation.}
\resizebox{0.95\linewidth}{!}{%
\begin{tabular}{lS[table-align-exponent=false]s[table-unit-alignment=left]ll}
Parameter&\multicolumn{2}{c}{Value}&Remarks&Source \\ \hline
$G_{APD}$&200&&& \\
$Y$&500&\per\mega\electronvolt&at \SI{-25}{\degreeCelsius}&\cite{emc-tdr} \\
$A_{rel}$&0.16&&&\cite{ds_apd}\\
$Q_{eff}$&0.7&&&\cite{ds_apd}\\
$e$&1.6e-19&\coulomb&& \\
$G_{PASA}$&2.2e11&\volt\per\coulomb&& \\ 
$n$&1.75&&& \\
$\tau_s$&280&\nano\second&& \\ \hline
\end{tabular}}
\end{table}

\begin{figure}[htb]
\includegraphics[width=\linewidth]{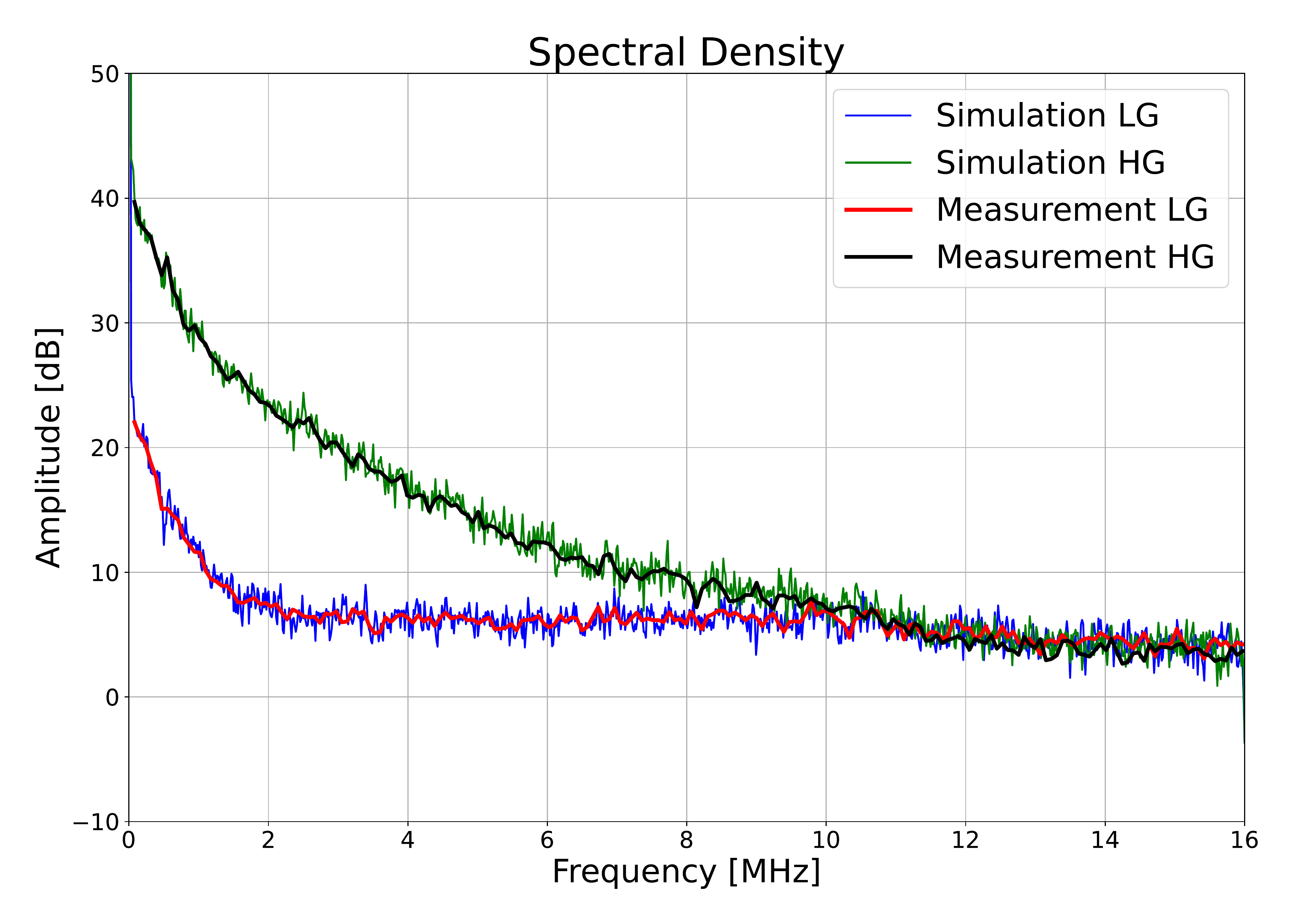}
\caption{\label{pic:NoiseSpec}Spectrum of the simulated noise in comparison with the measured noise of the APFEL ASIC.}
\end{figure}

Corresponding to the readout scheme with two APDs for each scintillation crystal and two outputs of the APFEL per APD with gain factors of 1 and 16 the numerical stimulus consisting of APFEL pulses is split into two branches and each branch is split into two channels with gain factors corresponding to the APFEL ASIC.\\
On each channel an individual noise signal is added with a noise spectrum that corresponds to measurements done with the APFEL ASIC. Figure \ref{pic:NoiseSpec} shows the noise spectra of the measured and the simulated APFEL noise.

For calibration a stimulus with cyclic pulses with constant amplitude corresponding to \SI{4800}{\mega\electronvolt} was used. The start times of the pulses were shifted in steps of 1/256 sampling periods to get an equally distributed fine time. \\

The numerical waveforms generated this way were used as stimulus for a full chip VHDL simulation of the ATR16 ASIC. The simulated output data were written into a file and analysed within the ROOT Framework\cite{BRUN199781,root-6-20-04}.\\

The first observation when analysing the data was that a comparison of the estimated times and amplitudes with fit parameters of fitted pulse functions shows a fine time dependency that has to be corrected.

\subsection{Fine time Correction}
\begin{figure}[htb]
\includegraphics[width=\linewidth]{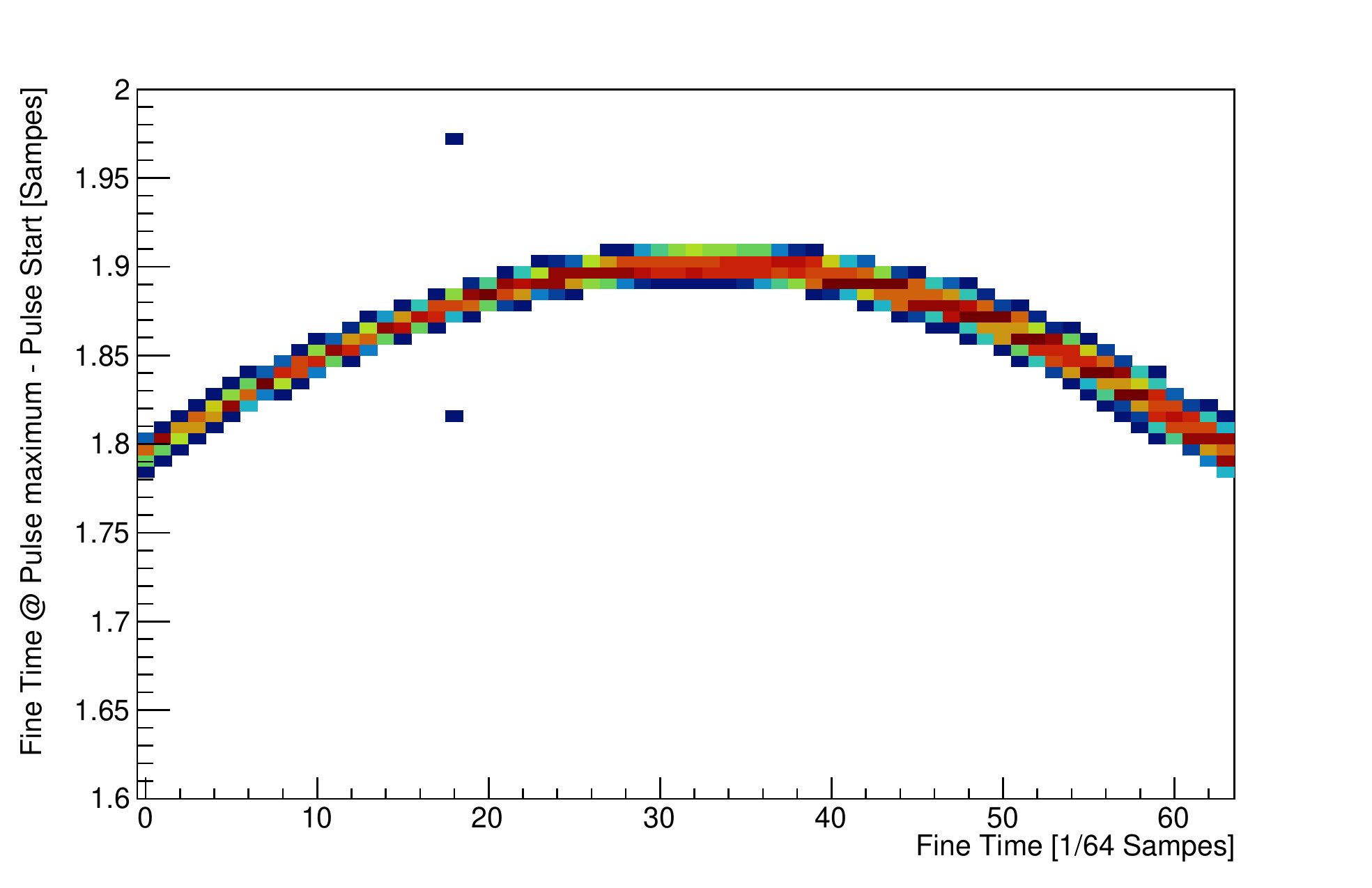}
\caption{\label{pic:timecharacteristics}Characteristics of the time of pulse maximum estimation algorithm as described in section \ref{sec:TimePulseMax}.}
\end{figure}

Figure \ref{pic:timecharacteristics} shows the observed fine time characteristics of the pulse maximum estimation algorithm as an example for the time estimation algorithms. On the x axis the estimated fine time in steps of 1/64 sampling intervals is plotted and on the y axis the difference between the estimated fine time and fine time determined with a pulse fit. As the time estimation algorithm uses in between the sampling points a linearisation of the non-linear pulse function an error is introduced which can be observed in the plot as fine time dependence. A similar behaviour is observable for the other algorithms.\\

To correct the observed non-linearities a look-up table is used. The correction vectors are generated with two different methods, the distribution analysis and the fit comparison. The distribution analysis requires equally distributed fine times and can be done with a data set containing estimated pulse features only. In difference, the fit comparison method does not require a special fine time distribution but raw data i.e. full pulse transients are needed to fit the analytic pulse function (equation \ref{eq_pasa_pulse}) to the transients.

\subsubsection{Non-Linearity Correction by Distribution Analysis}
The first method is a statistical method frequently used to correct non linearities of TDCs which was described by Pelka et. al. in \cite{irrDB142}. With a large number $N$ of calibration events the correction vector is mainly given by the integral non linearity error of the $j$th time bin

\begin{equation}
L_j = \sum\limits_{i=0}^j \left( \frac{n_i}{N} - \frac{1}{M} \right) 
\end{equation}

with the number of events $n_i$ in the $i$th fine time bin and the total number of fine time bins $M$.

\subsubsection{Non-Linearity Correction by Fit Parameters}
For the second method the difference between the estimated fine time and the fine time determined with the pulse fit is calculated and filled in histograms for each fine time bin. The correction value for a fine time bin is given by the mean value of the corresponding histogram.\\

\begin{figure}[htb]
\includegraphics[width=\linewidth]{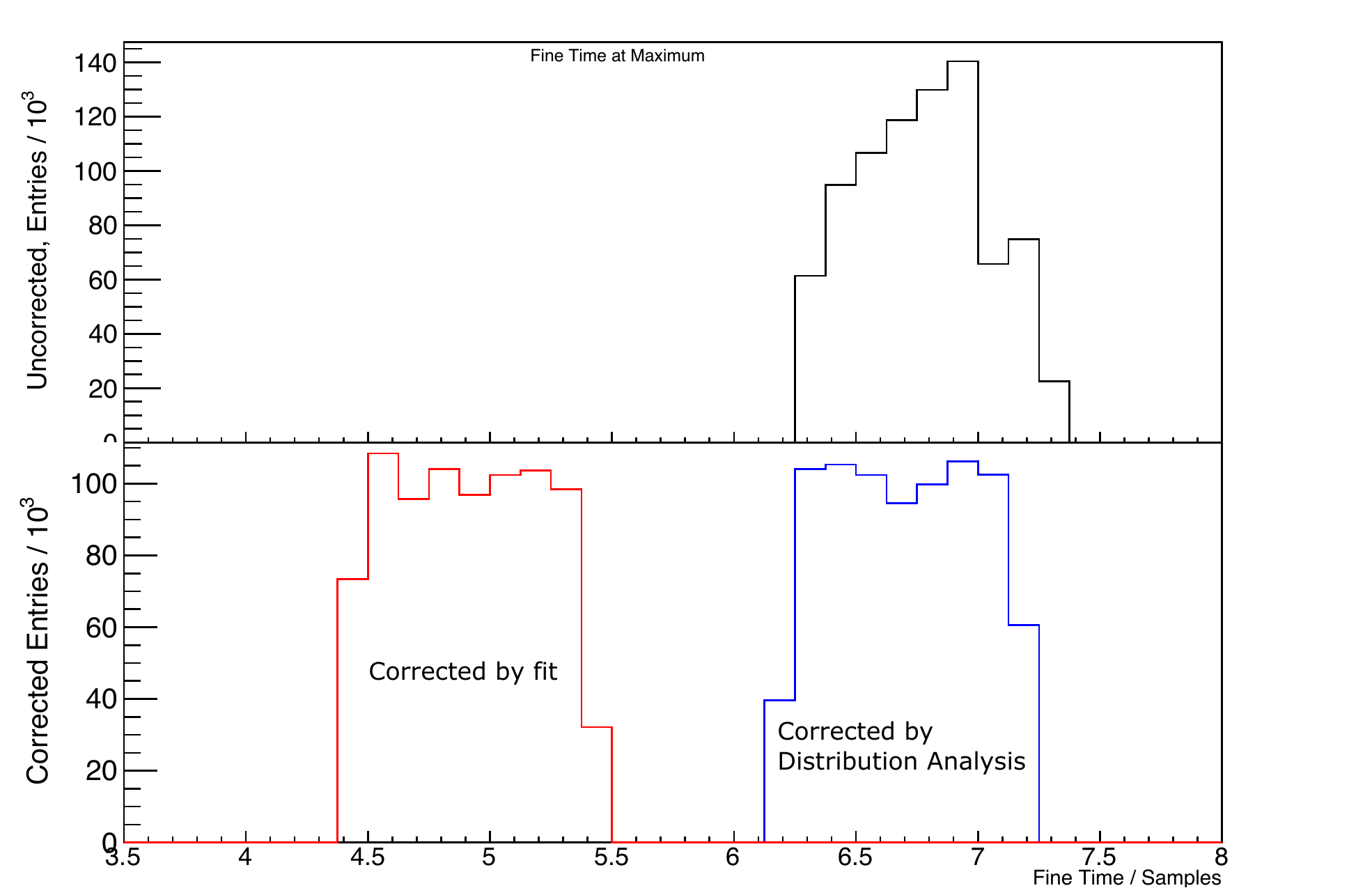}
\caption{\label{pic:timedist_correction}Distribution of estimated fine times without correction (top) and after correction (bottom) with the distribution analysis method and the fit method.}
\end{figure}
Figure \ref{pic:timedist_correction} shows the distribution of the estimated fine time with the pulse maximum algorithm without and with linearity correction. While the uncorrected distribution shows large structures the corrected distributions are almost flat. In addition to the linearity correction the fit method shifts the distribution as the pulse fit determines the time of the pulse start.

\subsection{Amplitude Correction}
\begin{figure}[htb]
\includegraphics[width=\linewidth]{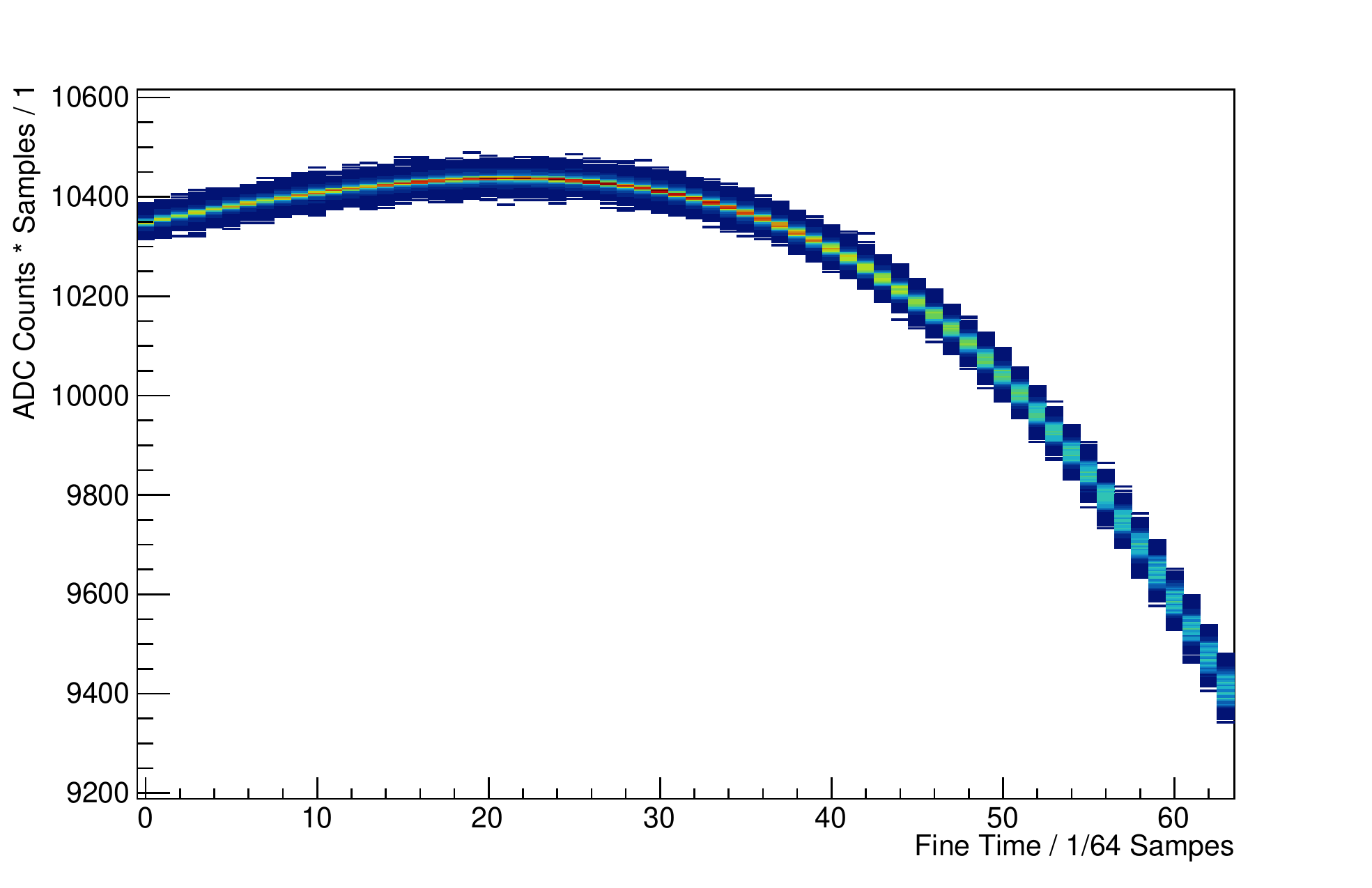}
\caption{\label{pic:winint_vs_finetime}Estimated amplitude with the window integral algorithm versus estimated fine time. For fine time estimation the pulse maximum algorithm described in section \ref{sec:TimePulseMax} is used.}
\end{figure}
Just as the estimated fine time the estimated amplitude shows a non linear fine time dependency. This is due to the fact that the value of the highest sampling point $a_m - \bar{b}$ is equal to the pulse amplitude $\hat{v}$ only in the case that the sampling coincides with the pulse maximum. In general $(a_m - \bar{b}) \le \hat{v}$ with the difference $\delta = \hat{v} - (a_m - \bar{b})$ which is a function of the pulse phase. This has an influence on the pulse maximum algorithm as well as on the integral algorithms and is shown for the window integral algorithm as an example in figure \ref{pic:winint_vs_finetime}.\\

To correct these dependencies vectors of correction factors are used. Again two methods are used to generate these correction vectors

\subsubsection{Correction by Scaling to Maximum Value}
In a first step, the fine time with the maximum estimated amplitude is searched. In the second step for each fine time the ratio between the estimated amplitude for this fine time and the maximum estimated amplitude is calculated as correction factor for this fine time.

\subsubsection{Correction by Fit}
For each fine time the ratio between the estimated amplitude and the amplitude of the fitted pulse is calculated and stored as correction factor.\\

\begin{figure}[htb]
\includegraphics[width=\linewidth]{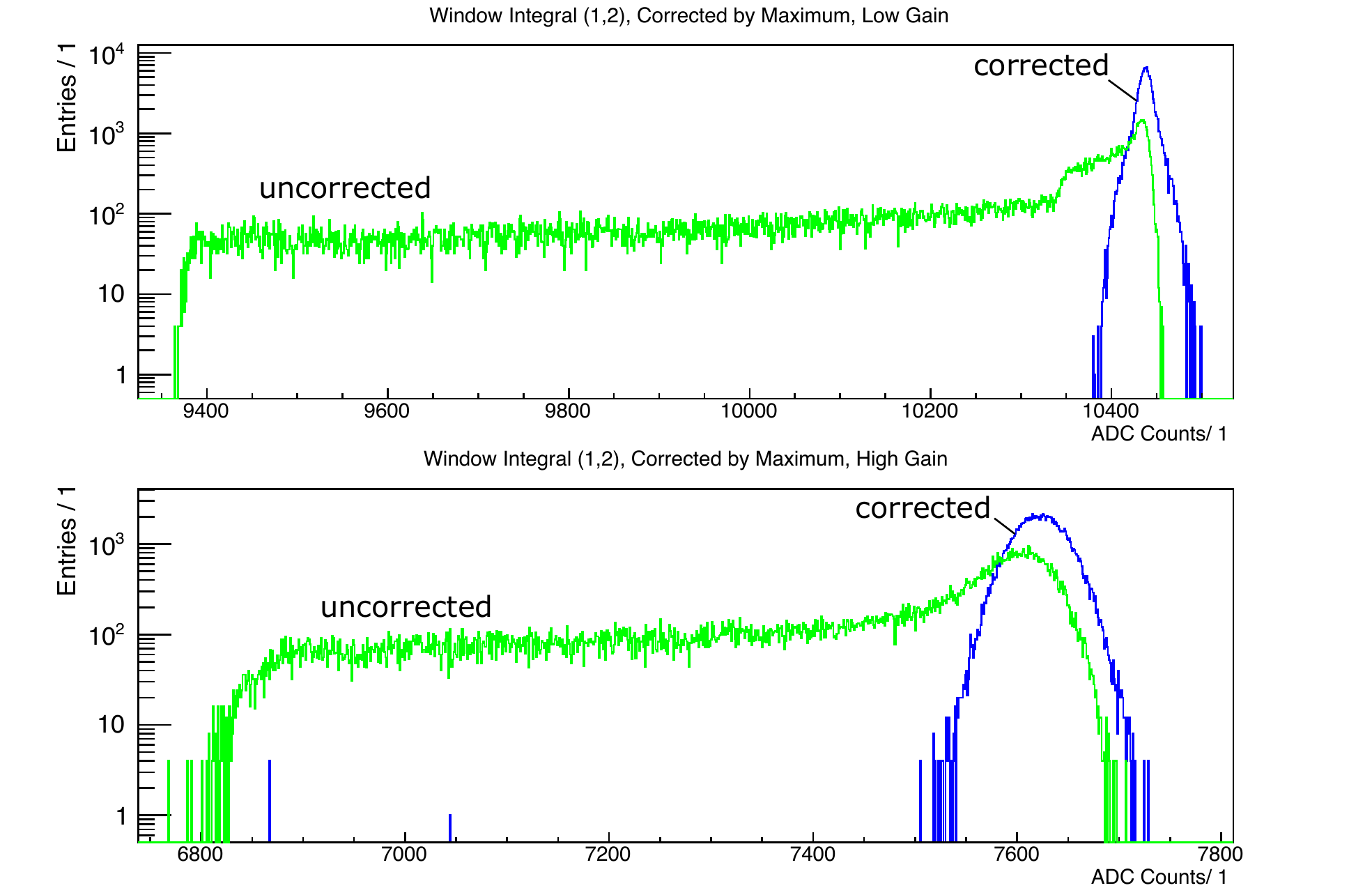}
\caption{\label{pic:AmpUncorrVsCorr}Amplitude spectrum for uncorrected estimated amplitudes and amplitudes corrected by scaling to maximum value.}
\end{figure}

Figure \ref{pic:AmpUncorrVsCorr} shows the amplitude spectrum of estimated amplitudes with the window integral algorithm without and with correction. 

\section{Analysis and Results}\label{sec:results}
The test data contain cyclic pulses with an interval of $T_{cyc} = \SI{4}{\micro\second} + T_S / 64$ with the sampling period $T_S = \SI{125}{\nano\second}$ and a geometric sequence of 15 pulse heights corresponding to energies from \SIrange{10}{4800}{\mega\electronvolt}. Cyclic pulses have been chosen to enable an analysis in absence of pile-up. An analysis of pile-up-effects will follow in section \ref{sec:pileup}.

\subsection{Time Estimation}

\begin{figure}[htb]
\includegraphics[width=\linewidth]{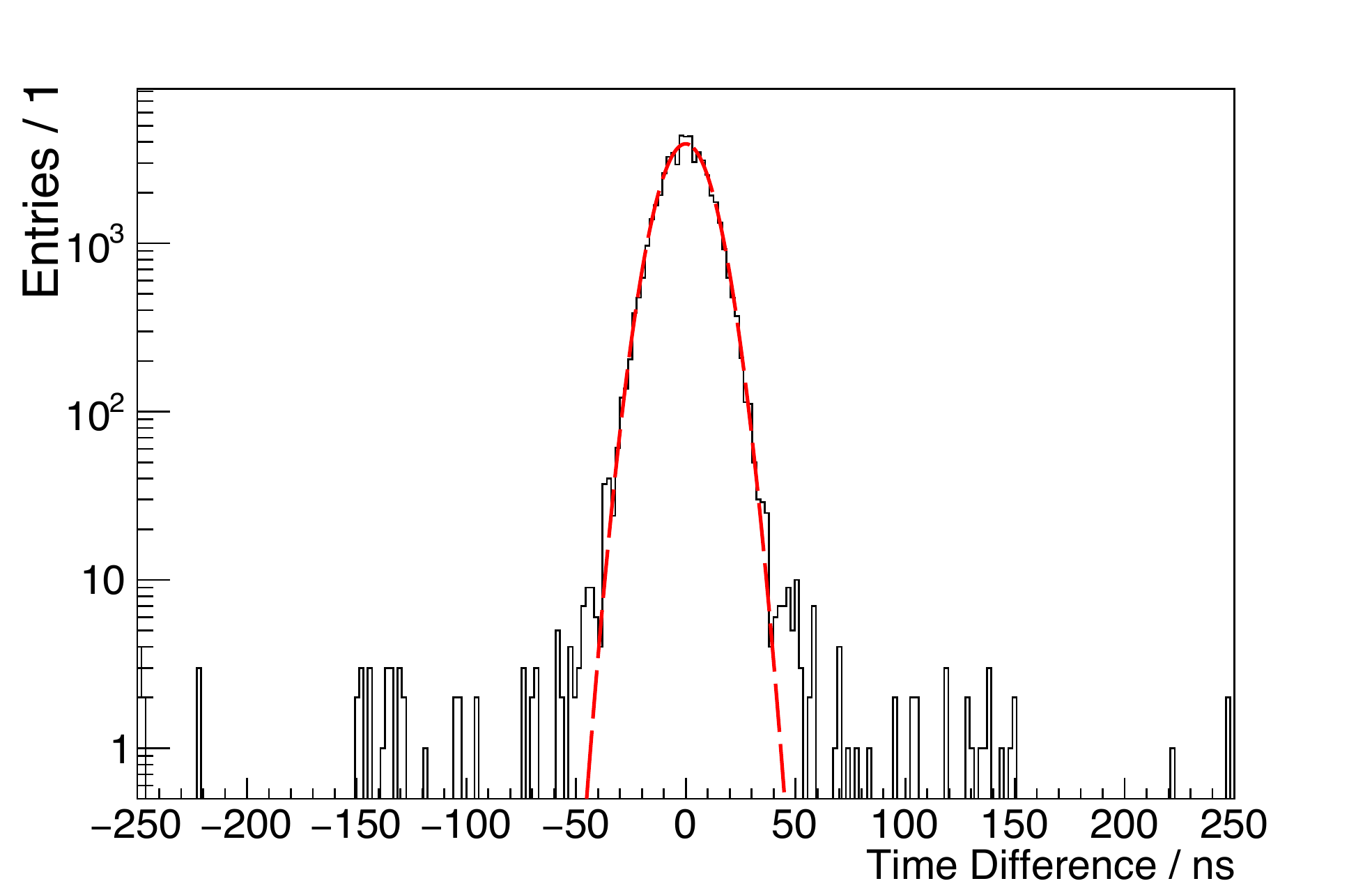}
\caption{\label{pic:Timedifference} Time difference between two channels fed with the same pulse and a Gaussian fit (dashed line). The data are estimated with the pulse maximum algorithm and corrected by distribution analysis.}
\end{figure}

To determine the precision of the estimated fine times a difference spectrum of the estimated values of two channels fed by the same pulse is generated each for the four estimation algorithms. Figure \ref{pic:Timedifference} shows the time difference spectrum for pulses with an energy of \SI{15.5}{\mega\electronvolt} in high gain channels estimated with the zero crossing of first derivative algorithm. A clear peak around zero is visible in the data which is well described by a Gauss function. \\

\begin{figure}[htb]
\includegraphics[width=\linewidth]{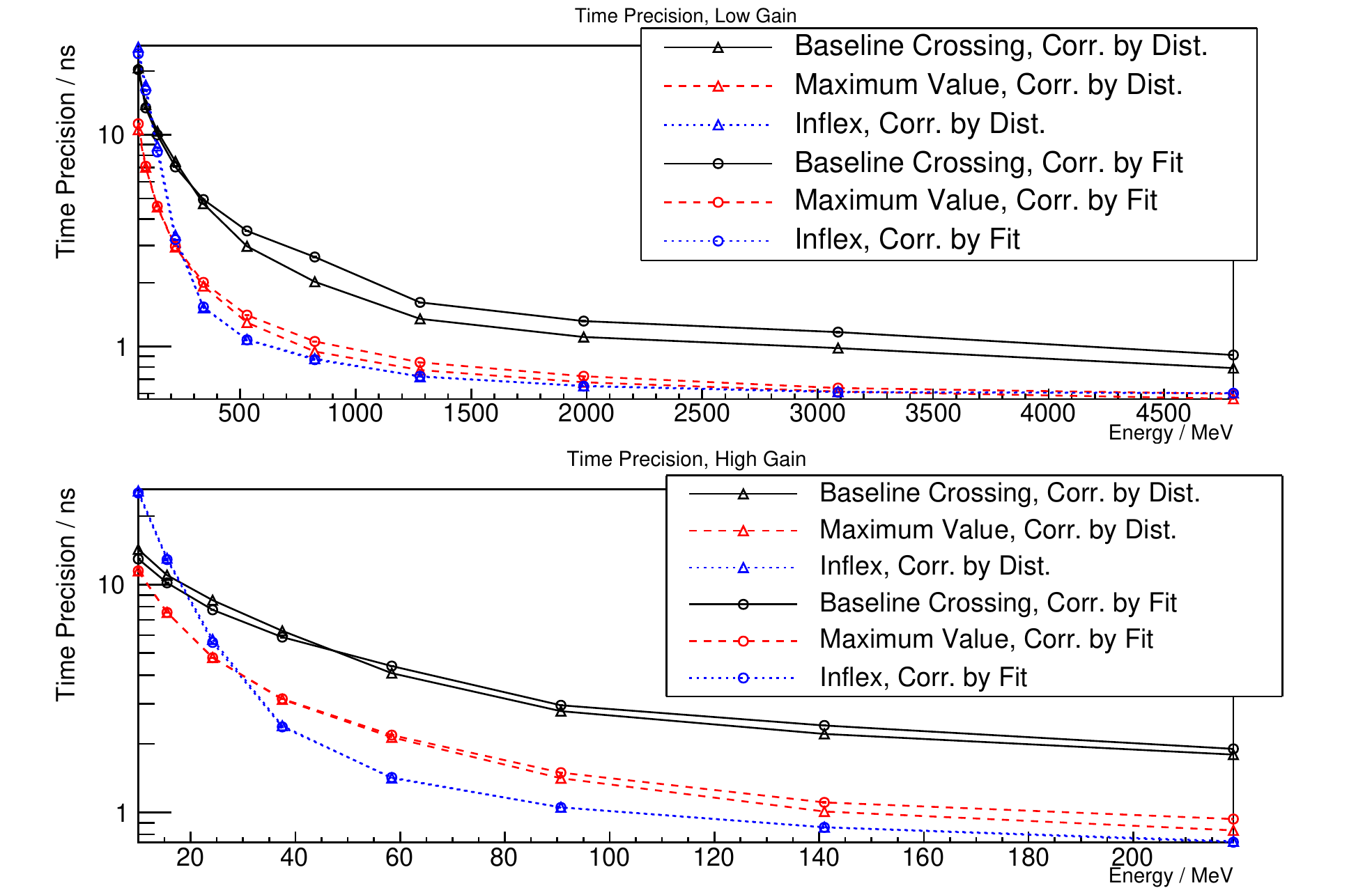}
\caption{\label{pic:Timeprec} Time precision in dependence of the pulse energy.}
\end{figure}

As the histogramed data are the difference of two channels $\sigma$ of the fitted Gauss function divided by square root of two is used as a measure of the time precision of a single channel. The time precision obtained this way is plotted for all estimation algorithms and pulse energies in low and high gain in figure \ref{pic:Timeprec}. The time precision gets better with increasing pulse energy. As listed in table \ref{tab:timeres} a time precision in the order of a nano second is obtained for the highest energy. 

\begin{table}[htb]
\caption{\label{tab:timeres}Obtained time precision for the highest energies in high and low gain mode. Data are corrected with the pulse fit method.}
\begin{center}
\begin{tabular}{lSSs}
&\multicolumn{1}{c}{High Gain}&\multicolumn{1}{c}{Low Gain}&\\ 
Estimation Algorithm&\multicolumn{1}{c}{\SI{219}{\mega\electronvolt}}&\multicolumn{1}{c}{\SI{4800}{\mega\electronvolt}}&\multicolumn{1}{c}{unit}\\ \hline
Baseline crossing&2.69&1.29&\nano\second \\
Time @ Maximum&1.32&0.84&\nano\second \\

\ifthenelse{\boolean{parabolic}}{%
Time @ Inflexion Point&2.52 &1.90 &\nano\second \\ 
Parabolic Approximation&1.28&0.79&\nano\second\\ \hline}{%
Time @ Inflexion Point&1.05 &0.85 &\nano\second \\ \hline
}
\end{tabular}
\end{center}
\end{table}

\subsection{Amplitude Estimation}
\begin{figure}[htb]
\includegraphics[width=\linewidth]{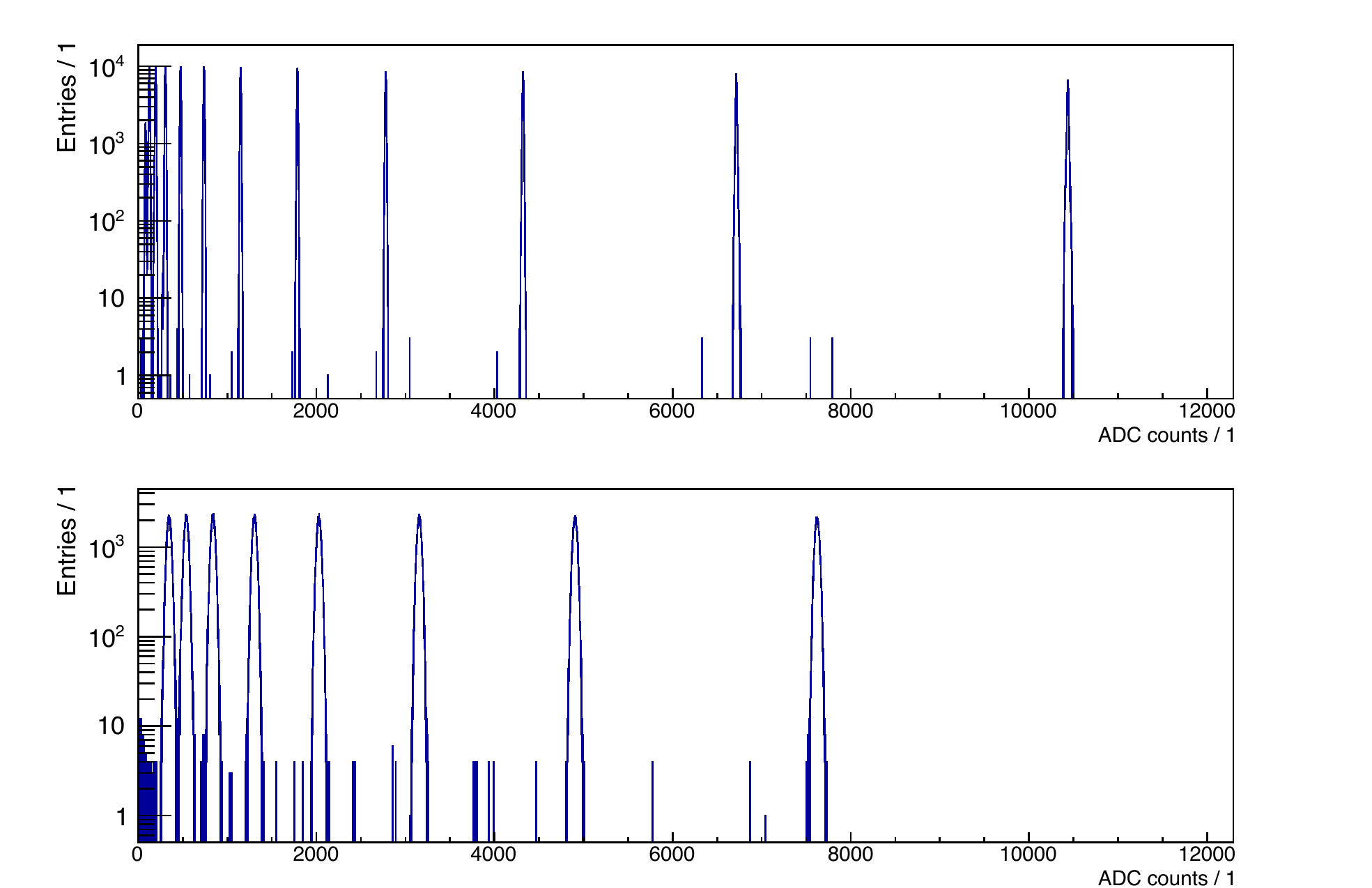}
\caption{\label{pic:EnergySpec} Pulse height spectrum. Data estimated with the window integral (1,2) algorithm and corrected with the scaling to maximum value method. Low gain branch in the upper and high gain branch in the lower histogram.}
\end{figure}
\begin{figure}[tb]
\includegraphics[width=\linewidth]{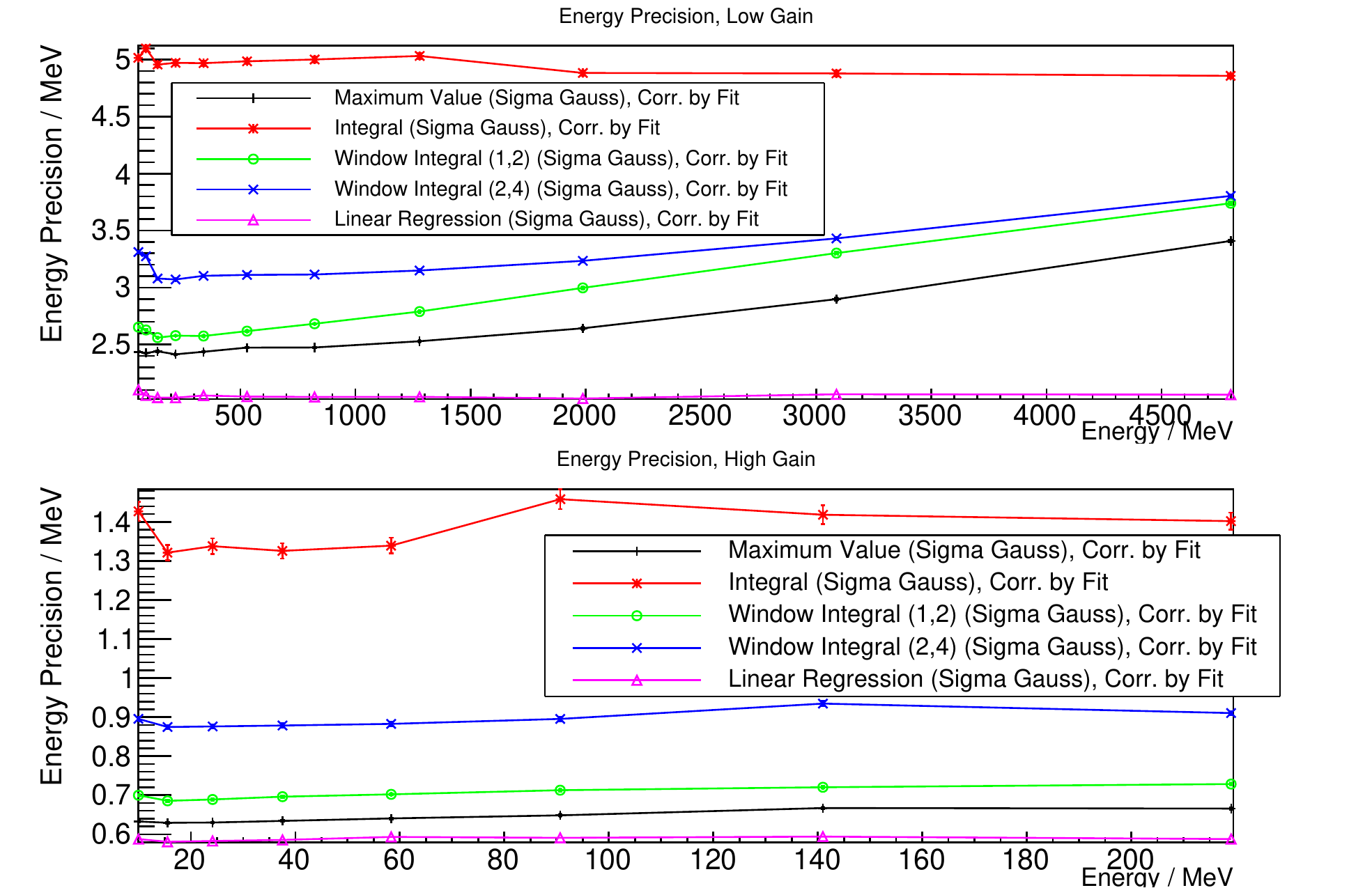}
\caption{\label{pic:EnergyPrec} Energy precision in dependency from pulse energy.}
\end{figure}

Figure \ref{pic:EnergySpec} shows the pulse height spectrum of pulses analysed with the window integral method. Equivalent spectra have been generated for all estimation algorithms. One can resolve 11 peaks in the low gain mode and 8 peaks in the high gain mode. These peaks are fitted with Gauss functions to get the mean value and the width of the peaks. From the correlation of the corresponding energy of each peak which is well known from the stimulus generation and the mean value of the Gauss functions a calibration factor can be evolved. So the energy precision is determined as product of this calibration factor and the sigma value of the Gauss function. The obtained energy precision is plotted in figure \ref{pic:EnergyPrec}.\\

\begin{figure*}
\begin{minipage}[b]{0.49\linewidth}
\makebox[0cm]{}\\
\includegraphics[width=\linewidth]{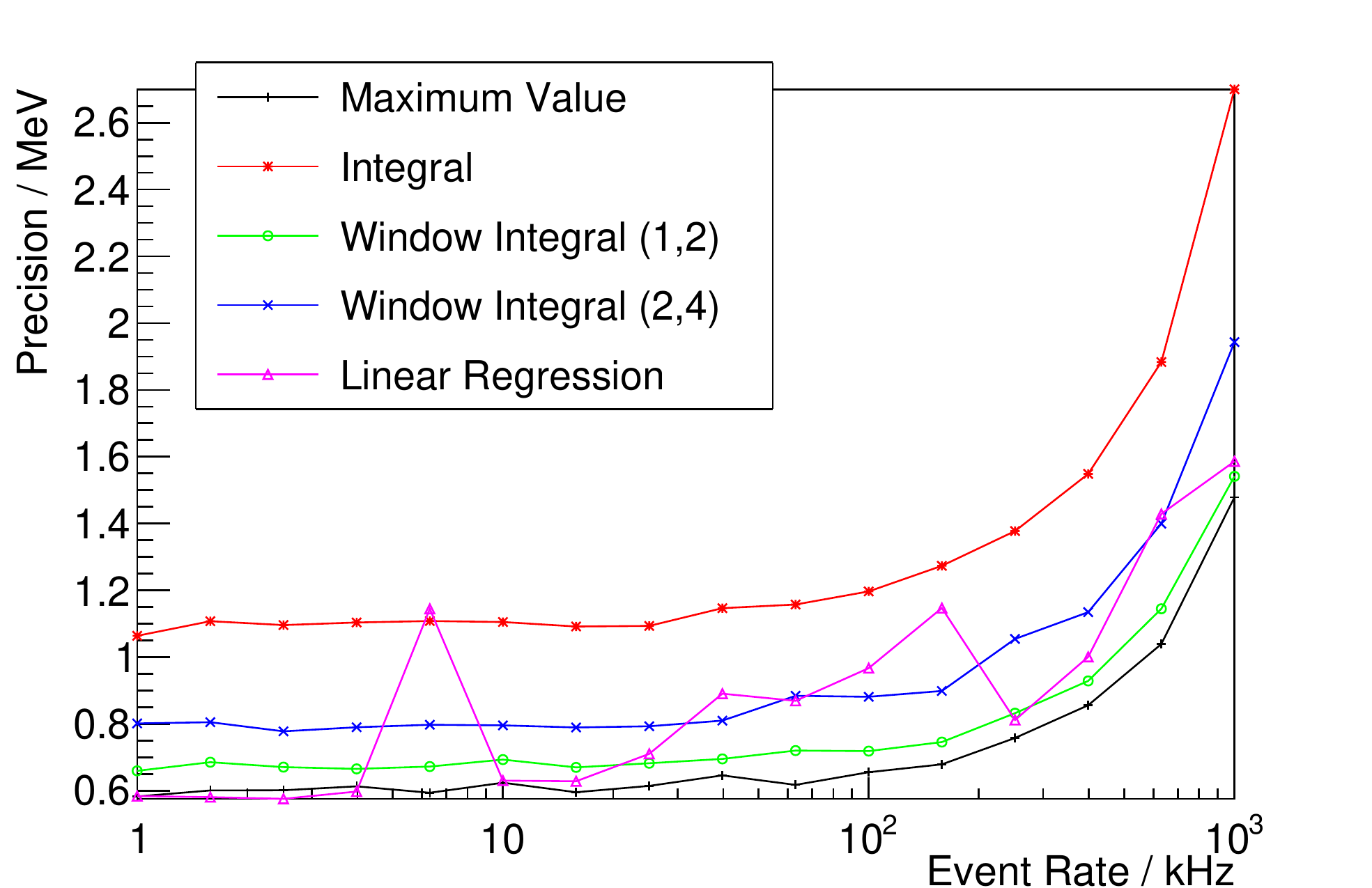}
\end{minipage}
\hfil
\begin{minipage}[b]{0.49\linewidth}
\makebox[0cm]{}\\
\includegraphics[width=\linewidth]{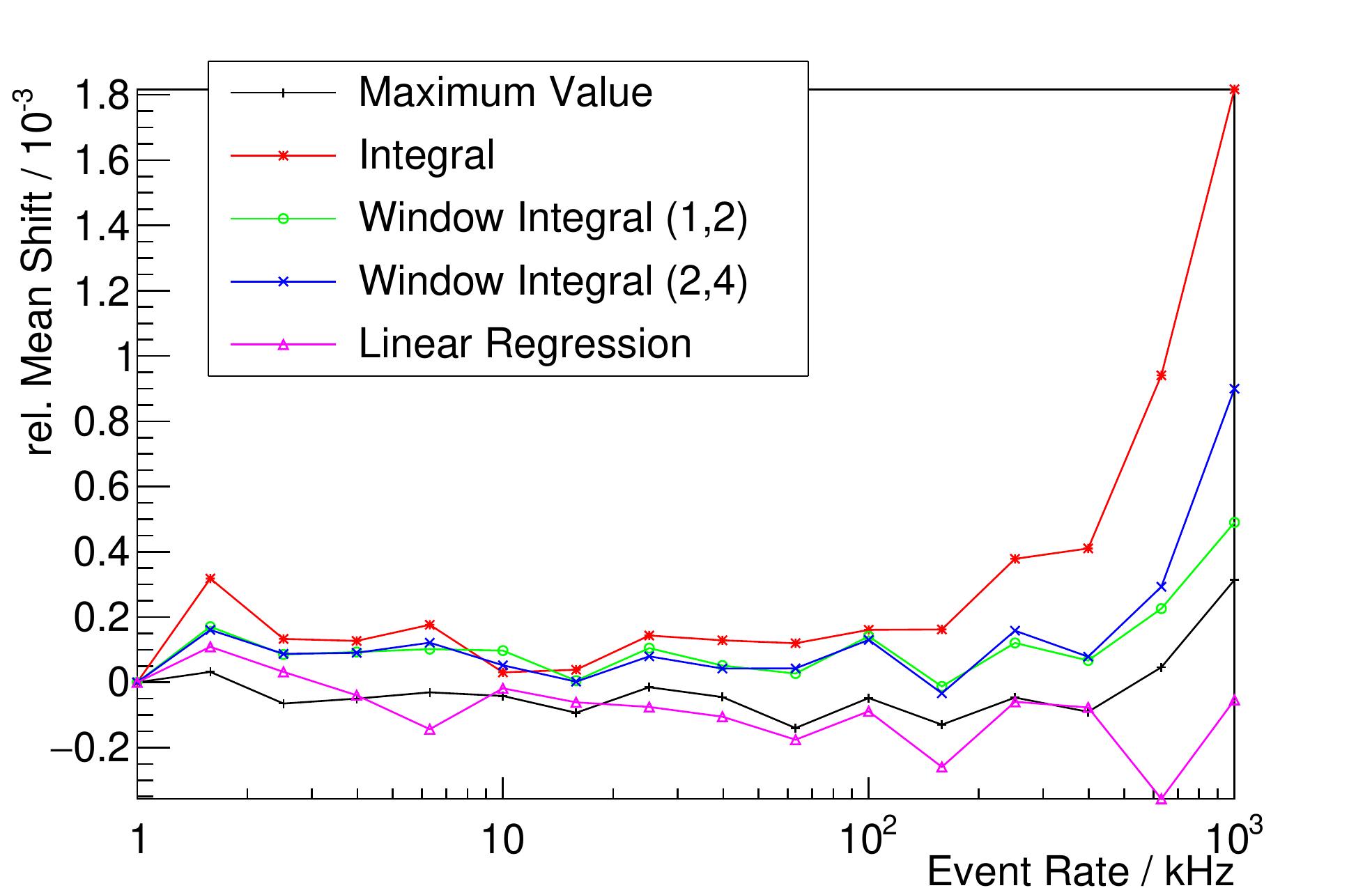}
\end{minipage}
\caption{\label{pic:MeanShiftAndStdDevVsRate}Standard deviation (left) and mean shift (right) as a function of the event rate for all events of an energy of \SI{219}{\mega\electronvolt} in the high gain mode.}
\end{figure*}

The energy precision by electronic noise and feature extraction shows no significant dependency from the pulse energy but the estimation performance of different algorithms varies by a factor of $\approx 2.5$.

\subsection{Effects of Pile-up}\label{sec:pileup}
\begin{figure}[tb]
\includegraphics[width=\linewidth]{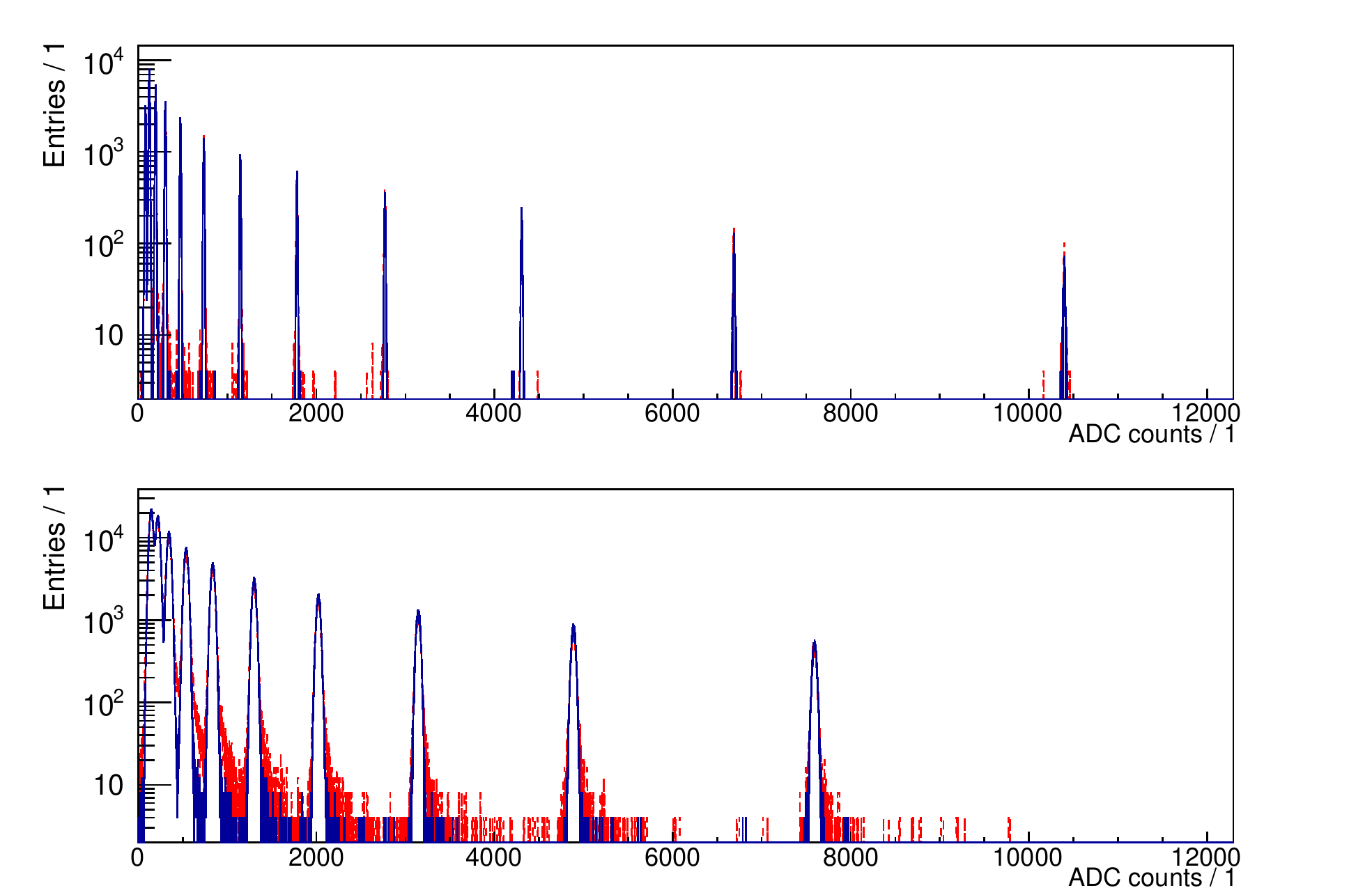}
\caption{\label{pic:EnergySpecPoisson} Pulse height spectrum of Poisson distributed pulses with \SI{251.2}{\kilo\hertz} pulse rate (red, dashed) and \SI{25.1}{\kilo\hertz} pulse rate (blue, solid).}
\end{figure}

The results of the previous sections are obtained by simulations with a cyclic stimulus that avoids pile-up events completely. To get a more realistic picture that includes the effects of pile-up the simulation and analysis was repeated with Poisson distributed stimuli with a geometric sequence of 16 event rates in the range from \SI{1}{\kilo\hertz} to \SI{1}{\mega\hertz}. The energy spectrum was extended with additional discrete energies down to \SI{500}{\kilo\electronvolt} and the incidences of the discrete energies are $1/E$ distributed.\\

To get sufficient statistics for highest energies at lowest rates in a reasonable time it was abstained from performing a full chip VHDL simulation instead, a chip model was developed in C++.\\

The estimation algorithm checks the traces for a second increasing on the pulse tail to detect pile-up events. Those events are rejected and the following analysis is based on events with no or no obvious pile-up.\\

Figure \ref{pic:EnergySpecPoisson}  shows the same spectrum for Poisson distributed pulses as shown in figure \ref{pic:EnergySpec} for cyclic pulses. The most obvious difference is that the spectra for Poisson distributed pulses become asymmetric with high energy tails. Comparing the spectra for \SI{25.1}{\kilo\hertz} rate and \SI{251.2}{\kilo\hertz} one gets the impression that the asymmetry increases with increasing rate.
A similar behaviour was also described for high rate detector measurements with a sampling ADC readout in \cite{OliverNoll}. \\

To quantify the effect of the high energy tails in the energy spectrum the mean value as well as the standard deviation of the estimated energy of all events in defined ranges is calculated. The boundaries of these ranges are set to the geometric mean between two peaks. Figure \ref{pic:MeanShiftAndStdDevVsRate} shows the standard deviation and the relative shift of the mean value calculated for the energy range around \SI{219}{\mega\electronvolt} in the high gain mode.\\

The standard deviation increases for event rates above\linebreak \SI{100}{\kilo\hertz}. Up to \SI{500}{\kilo\hertz} a precision of \SI{1}{\mega\electronvolt} can be obtained by the window integral (1,2), \ifthenelse{\boolean{parabolic}}{maximum value and parabolic approximation}{and maximum value} algorithms. The relative shift of the mean value for these algorithms is below $\pm1\permil$. The irregular behaviour of the standard deviation of the linear regression estimation algorithm is not yet understood, but as only a minor benefit in precision could be obtained with this algorithm while it requires a large numerical effort it is not considered for practical use anyway.\\

\begin{figure}[htb]
\includegraphics[width=\linewidth]{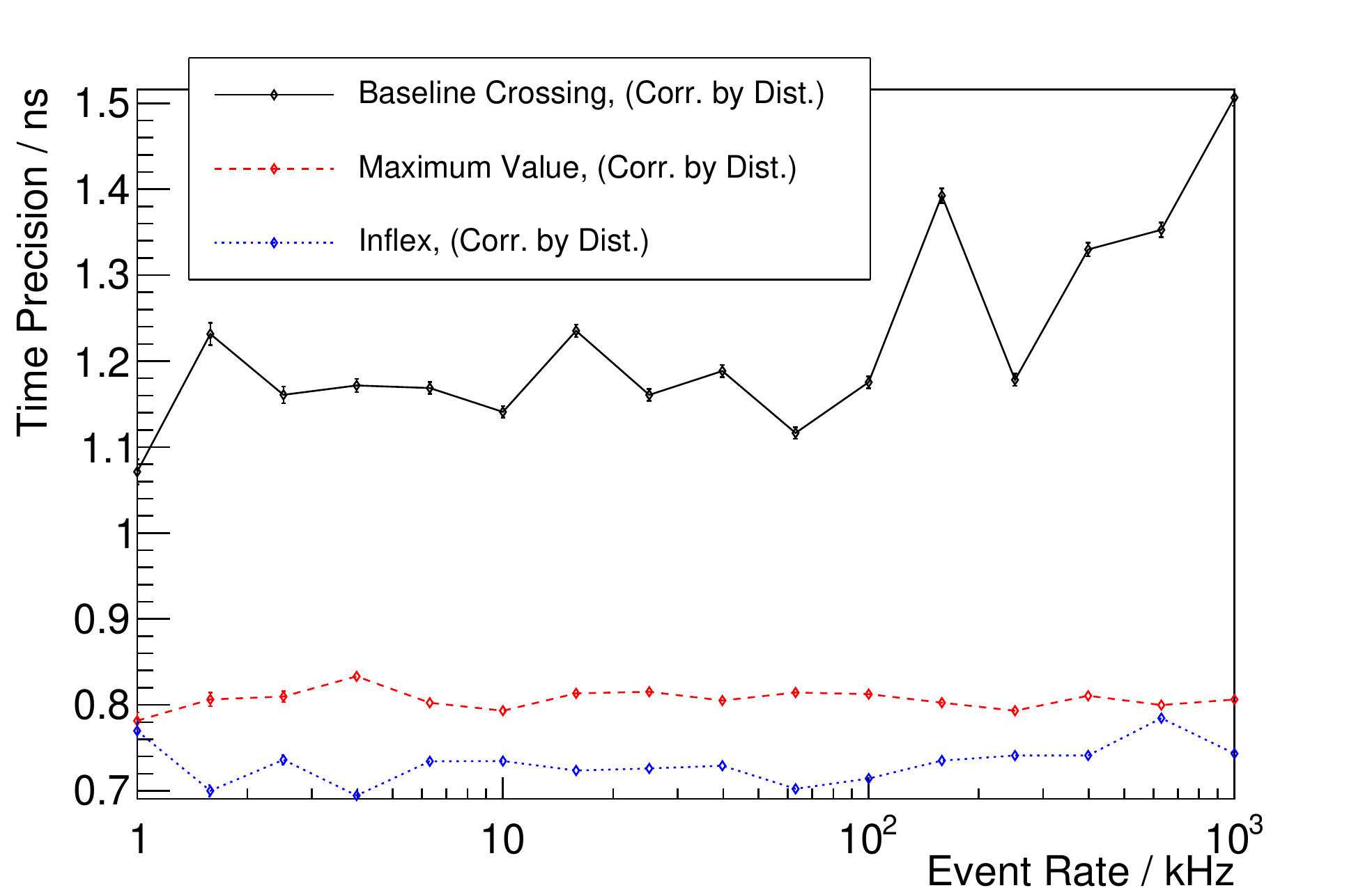}
\caption{\label{pic:TimePrec_vs_rate} Time precision at \SI{219}{\mega\electronvolt} in high gain mode as a function of event rate for Poisson distributed events.}
\end{figure}

The time precision is not affected by pile-up as in figure \ref{pic:TimePrec_vs_rate} no rate dependency is observable. 

\section{Summary and Conclusion}\label{sec:conclusion}
For this work simulated pulse transients of 16 samples have been analysed with \ifthenelse{\boolean{parabolic}}{four algorithms for time estimation and five algorithms for amplitude estimation}{three algorithms for time estimation and four algorithms for amplitude estimation}. The algorithms and corrections have been shown. For amplitude estimation the most elaborate algorithm (linear regression, see section \ref{sec:linregression}) provides the best energy precision as expected but the precision of the very simple maximum value algorithm (section \ref{sec:AmpPulseMax}) is only slightly worse. Surprisingly the energy precision does not benefit from integration across several sampling bins. A reason for this might be that the noise is dominated by low frequency components while integration suppresses high frequency noise only.\\

As the algorithms that obtain the best amplitude precision require only a few samples around the pulse maximum the total number of recorded samples could even be reduced. Nevertheless it was decided to keep 16 sample traces in favour of flexibility and compatibility with other front ends which are planned to be connected to the same recording backend\cite{twepp21-TR,twepp21-AWAGS}.\\ 

The time precision gets better with increasing pulse amplitude. Even though the sampling period of the transients is \SI{125}{\nano\second} a time precision in the order of \SI{1}{\nano\second} can be obtained.\\

In addition the rate dependency of realistic Poisson distributed events was analysed. With increasing event rate pile-up events lead to a high energy tail and a slightly degrading energy precision that still fulfils the requirements while the time precision does not depend on the event rate.\\

The results finally show that feature extraction of short event transients is feasible with a sufficient energy and time precision. Based on this analysis the development of the ATR16 was continued and an on-chip feature extraction unit was implemented realising the algorithms investigated in this work.


\section*{Funding}
This research did not receive any specific grant from funding agencies in the public, commercial, or not-for-profit sectors.

\section*{Author Contribution}
{\bf Holger Flemming}: Conzeptualisation, Methodology, Software, Formal analysis, writing - Original  draft, {\bf Oliver Noll}: Software, Formal analysis, Writing - Review and Editing

\bibliographystyle{elsarticle-num}
\bibliography{../biblio/ASIC_Design}

\end{document}